\documentclass[fleqn,usenatbib]{mnras}
\usepackage{graphicx}
\usepackage{txfonts}

\usepackage{url}
\usepackage{graphicx}
\usepackage[hyphenbreaks]{breakurl}
\usepackage{float}
\usepackage{subfig}
\usepackage{multirow}

\title[TESS observations of low mass rotators]{{\sl TESS} observations of southern ultra fast rotating low mass stars}

\author[]
{Gavin Ramsay,$^{1}$ J. Gerry Doyle, $^{1}$ Lauren Doyle,$^{1,2}$\\
$^{1}$Armagh Observatory \& Planetarium, College Hill, Armagh, BT61 9DG, UK\\
$^{2}$Mathematics, Physics and Electrical Engineering, Northumbria University, Newcastle upon Tyne, NE1 8ST, UK\\}

\date{Accepted 2016 June 23. Received 2016 June 23; in original form 2016 April 5}

\begin{document}
\outer\def\gtae {$\buildrel {\lower3pt\hbox{$>$}} \over 
{\lower2pt\hbox{$\sim$}} $}
\outer\def\ltae {$\buildrel {\lower3pt\hbox{$<$}} \over 
{\lower2pt\hbox{$\sim$}} $}
\newcommand{\Msun}{$M_{\odot}$}
\newcommand{\lsun}{$L_{\odot}$}
\newcommand{\Rsun}{$R_{\odot}$}
\newcommand{\solar}{${\odot}$}
\newcommand{\kep}{\sl Kepler}
\newcommand{\ktwo}{\sl K2}
\newcommand{\tess}{\sl TESS}
\newcommand{\swift}{\it Swift}
\newcommand{\Porb}{P_{\rm orb}}
\newcommand{\nuorb}{\nu_{\rm orb}}
\newcommand{\eplus}{\epsilon_+}
\newcommand{\eminus}{\epsilon_-}
\newcommand{\cd}{{\rm\ c\ d^{-1}}}
\newcommand{\MdotL}{\dot M_{\rm L1}}
\newcommand{\Mdot}{$\dot M$}
\newcommand{\Mdotsolar}{\dot{M_{\odot}} yr$^{-1}$}
\newcommand{\Ldisk}{L_{\rm disk}}
\newcommand{\src}{KIC 9202990}
\newcommand{\ergscm} {erg s$^{-1}$ cm$^{-2}$}
\newcommand{\rchi}{$\chi^{2}_{\nu}$}
\newcommand{\chisq}{$\chi^{2}$}
\newcommand{\pcmsq} {cm$^{-2}$}

\providecommand{\lum}{\ensuremath{{\cal L}}}
\providecommand{\mg}{\ensuremath{M_{\rm G}}}
\providecommand{\bcg}{\ensuremath{BC_{\rm G}}}
\providecommand{\mbolsun}{\ensuremath{M_{{\rm bol}{\odot}}}}
\providecommand{\teff}{\ensuremath{T_{\rm eff}}}

\maketitle
\begin{abstract}
In our previous study of low mass stars using {\tess}, we found a handful
which show a periodic modulation on a period $<$1 d but also displayed no
flaring activity. Here we present the results of a systematic search
for Ultra Fast Rotators (UFRs) in the southern ecliptic hemisphere
which were observed in 2 min cadence with {\tess}. Using data
from Gaia DR2, we obtain a sample of over 13,000 stars close to the
lower main sequence. Of these, we identify 609 stars which lie on the
lower main sequence and have a periodic modulation $<$1 d. The
fraction of stars which show flares appears to drop significantly at
periods $<$0.2 d. If the periods are a signature of the rotation
rate, this would be a surprise, since faster rotators would be expected
to have a stronger magnetic field and, therefore, produce more
flares. We explore possible reasons for our finding: the flare
inactive stars are members of binaries, in which case the stars rotation rate
could have increased as the binary orbital separation reduced due to
angular momentum loss over time, or that enhanced emission occurs at
blue wavelengths beyond the pass band of {\tess}. Follow-up
spectroscopy and flare monitoring at blue/ultraviolet wavelengths of
these flare inactive stars are required to resolve this question.
\end{abstract}


\section{Introduction}

Ultra Fast Rotators (UFRs) were first identified in the Pleiades open
cluster through photometric and spectroscopic observations
\citep[see][]{VanLeeuwen1982}. Further observations, such as outlined in
\citet{Stauffer1984}, indicated that these UFRs were heavily spotted
stars which had only recently arrived on the main sequence.  Large
starspots imply the presence of a strong magnetic field, which has
likely been generated through the dynamo mechanism
\citep[see][]{Hartmann1987,Maggio1987}, therefore, they are expected to be
active stars. Models of how low mass stars reach the main sequence
indicate that UFRs can be made if the magnetic field of the star
saturates at a critical rotational velocity \citep{BarnesSofia96}.

A key element in the study of stellar magnetism is that of stellar
flares, which have been observed across all wavelengths with
amplitudes of up to nine magnitudes being observed in the optical
\citep[e.g.][]{Simonian2014}. Although flares have been seen from
stars with earlier spectral-types, they appear more common in low mass
stars, with more flares being observed from stars which are fully
convective \citep[e.g.][]{Pettersen1989}. We have been making a study
of low mass stars using {\sl K2} \citep[][Paper I]{Doyle2018} and {\sl
  TESS} \citep[][Paper II]{Doyle2019} looking in particular at whether
the occurrence of flares is correlated with the stars rotation phase.

Until recently, determining the rotation period of stars was a
laborious and time consuming process. Matters changed with the launch
of {\sl Kepler} in 2009 which, observed the same 115 square degree
patch of sky just above the northern Galactic plane between Cygnus and
Lyra for four years \citep{Borucki2010}. Although its prime goal was
to discover planets around other stars, it provided the data to
measure the rotation period of tens of thousands of stars along the
main sequence \citep[e.g.][]{McQuillan2014}. After the initial four
year mission, {\sl Kepler} made a series of observations of fields
along the ecliptic plane each lasting several months, with the mission
being re-named {\sl K2}. Studies of open clusters of different ages
were able to shed further light on the stellar rotation period as a
function of age \citep[e.g.][]{Rebull2016}. Towards the end of the
{\sl K2} mission, {\sl TESS} was launched in 2018 and is currently
making a near all-sky survey with each 'sector' being observed for
approximately one month \citep{Ricker2015}.

During our studies of low mass stars made using {\sl TESS} (Paper II),
we identified a number of M dwarf UFRs possessing a
rotation period $<$0.3 d. Surprisingly, some of the UFRs we
identified show no or few flares in their lightcurves. Given that fast
rotation is generally taken to imply a strong magnetic field
\citep[e.g.][]{Parker1979}, this is unexpected.

In our previous studies we have targeted stars observed with {\sl K2}
and {\sl TESS} which had a known spectral type. In this study, we have
used data from Gaia DR2 to obtain a `blind' search for low mass stars
which have a rotation period $<$0.3 d. We then determine how many of
these UFRs show flares, determining their energy and comparing their
characteristics with stars with longer rotation periods.

\section{The TESS sample}
\label{tess}

In our previous studies, we selected targets which had 1 min cadence
{\sl K2} observations (Paper I) and 2 min cadence {\sl TESS}
observations (Paper II) of low mass stars which had a spectral type
later than M0 in the {\tt SIMBAD}
catalogue\footnote{\url{http://simbad.u-strasbg.fr/simbad/}}. Given
the fraction of low mass stars with spectral types determined by
spectroscopic observations is very far from complete, and that the
luminosity class can be incorrect, here we use the Gaia DR2 catalogue
\citep{GaiaBrown2018} to identify UFRs which have been observed using
      {\sl TESS} in 2 min cadence mode.

The {\sl TESS} satellite was launched on the 18th April 2018 into an
orbit with a period of 13.7 d. It has four small telescopes that cover
a 24$^{\circ}\times90^{\circ}$ instantaneous strip of sky
\citep[see][for details]{Ricker2015}. In comparison to {\sl Kepler},
      {\tess} is best suited to observe stars which are bright and
      nearby, but has an instantaneous sky coverage of 20 times
      greater. In contrast to {\sl Kepler}, {\tess} stays on a single
      field for 28 d, and in contrast to {\sl K2}, it will survey the
      entire sky apart from a few degrees close to the ecliptic
      equator over the initial two year mission. In addition, {\tess}
      has a continuous viewing zone centered on the ecliptic poles
      where stars can be observed in all 13 sectors, with observations
      totalling approximately one year. Each full-frame image has an
      exposure of 30 min which is downloaded to Earth and made
      publicly available. However, for around 20,000 stars in each
      sector, photometry with a cadence of 2 min is obtained, with
      most targets being selected from the community via a call for
      proposals.

In this study we have taken the catalogues of stars in Sectors
1--13 \footnote{\url{https://tess.mit.edu/observations}} which have 2
min photometry. This gives a total of 247,899 lightcurves, see Table
\ref{table1} for a detailed account of how many stars were left at
each stage of the selection process. There were 128,292 unique stars
with 2 min cadence lightcurves. For these stars, we matched them with
the Gaia DR2 catalogue \citep{GaiaBrown2018} using a cross match of
radius 4$^{''}$.

We take the parallax from the {\sl Gaia} DR2 catalogue and infer the
distance following the guidelines of \citet{BailerJones2015,Astra2016}
and \citet{GaiaLuri2018}, which is based on a Bayesian approach. In
practise we use a routine in the {\tt STILTS} package
\citep{Taylor2006} and use a scale length L=1.35 kpc, which is
appropriate for stellar populations in the Milky Way in general. We
use this distance to determine the absolute magnitude in the Gaia $G$
band (a very broad optical filter), $M_{G}$, using the mean Gaia $G$
magnitude. In our final sample, the ratio of the parallax to its error
is typically several hundred and we are, therefore, confident that the
resulting distances are robust, with 99 percent of our sample lying
within 220 pc. The other key observable is the blue ($BP$) and the red
($RP$) filtered magnitudes, which are derived from the Gaia Prism
data.

\begin{figure}
    \centering
    \includegraphics[width = 0.45\textwidth]{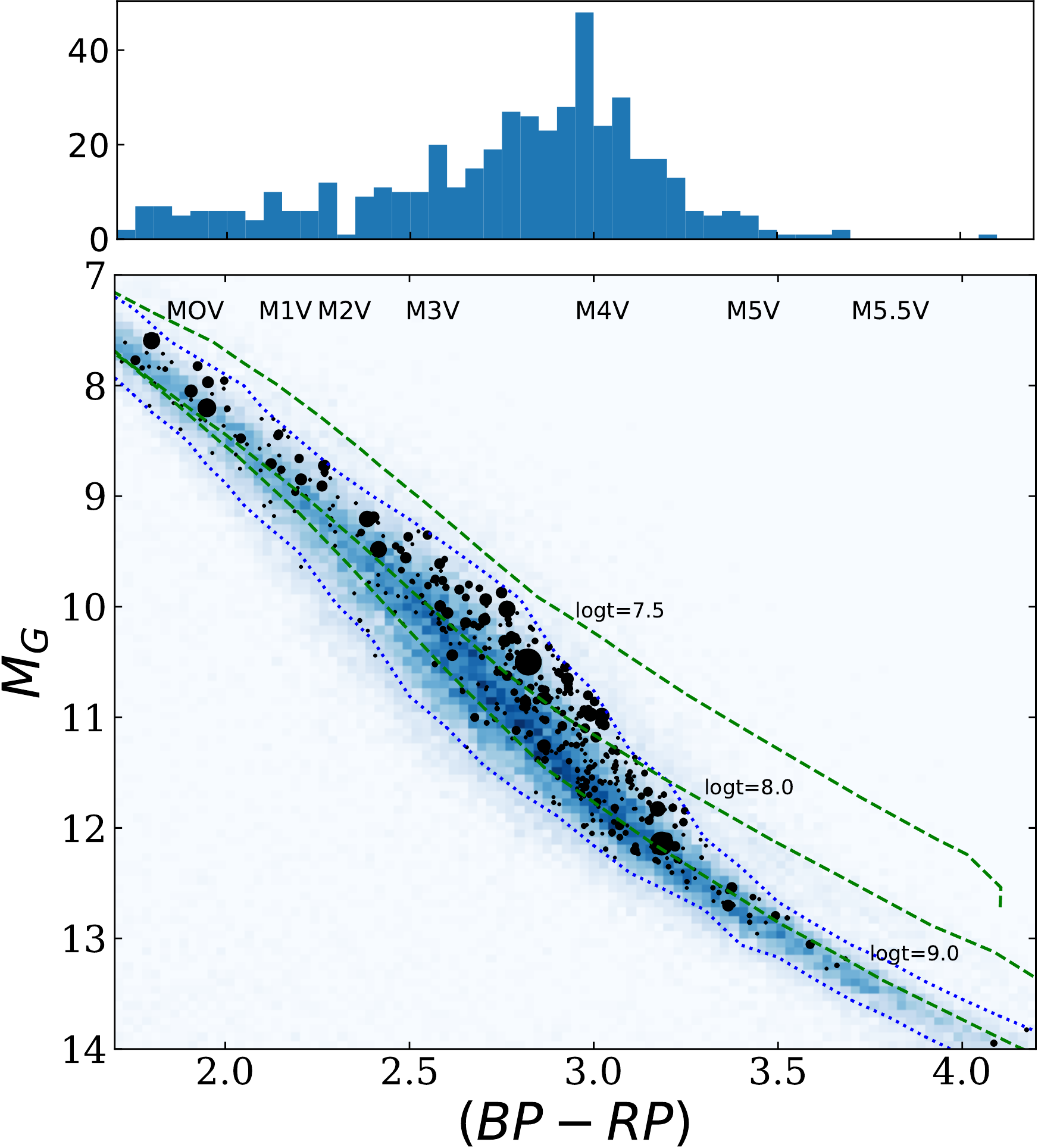}
    \caption{The colour-absolute magnitude diagram of stars within 150
      pc obtained using Gaia DR2 data, shown as a density in blue in
      the background. Stars which we have determined a rotation period
      $<$1 d are shown as black circles whose size is proportional to
      the flare rate (the smallest dots indicate no flares were
      detected). The blue dotted lines around either side of the main
      sequence track indicate the area where we selected stars for our
      sample. Age tracks are shown as dashed green lines. We show the
      blue ($BP$) and the red ($RP$) colours of spectral sub-types
      taken from the work of Eric Mamajek\protect\footnotemark.  In
      the top panel a histogram shows the number of stars over the
      range in $(BP-RP)$}.
    \label{gaia-hrd}
\end{figure}

Our aim is to identify low mass stars which are close to the main
sequence. We show the density of stars within 150 pc in the Gaia
$(BP-RP)$, $M_{G}$ plane in Figure \ref{gaia-hrd} where the lower main
sequence is well defined. We, therefore, select a region shown by
dotted lines in Figure \ref{gaia-hrd} to select stars observed using
{\sl TESS} in 2 min cadence mode, overall, there are 13,836 stars
within this region. They have ($BP-RP)$ colours which imply they have
spectral types in the range K9V to M6V, with there being an
enhancement of stars near M4V. We also show three age tracks which are
based on the {\tt PARSEC} evolutionary tracks \citep{Bressan2012} and
obtained from the Osservatorio Astronomico di Padova
portal\footnote{\url{http://stev.oapd.inaf.it/cgi-bin/cmd_3.3}} where
we have assumed default input parameters. These tracks indicate all
our stars are older than at least 30 Myr, with the redder stars having
ages of 1 Gyr or more.

\footnotetext{\url{http://www.pas.rochester.edu/~emamajek/EEM\_dwarf\_UBVIJHK\_colors\_Teff.txt}}

\section{Data analysis}

We downloaded the calibrated lightcurves of all 13,836 stars from
Sectors 1--13 from the MAST data
archive\footnote{\url{https://archive.stsci.edu/tess/}}. We used the
data values for {\tt PDCSAP\_FLUX}, which are the Simple Aperture
Photometry values, {\tt SAP\_FLUX}, after the removal of systematic
trends common to all stars in that Chip. Each photometric point is
assigned a {\tt QUALITY} flag which indicates if the data has been
compromised to some degree by instrumental effects. We removed those
points which did not have {\tt QUALITY=0} and normalised each light
curve by dividing the flux of each point by the mean flux of the
star. For those stars with data from more than one sector we made one
lightcurve by combining the sectors.

For each lightcurve we used the Lomb Scargle (LS) Periodogram as
implemented in the {\tt VARTOOLS} suite of tools to determine whether
a star showed significantly periodic modulation
\citep{HartmannBakos2016}. Since the lightcurves can still have
systematic trends present even after a global trend has been removed,
choosing the limit for identifying stars which show a significant
period is not clear cut. After manual inspection of a subset of light
curves and LS power spectra, we selected stars which had a False Alarm
Probability (FAP) such that log$_{10}$(FAP)$<$-50. All lightcurves
which were selected were then subject to manual inspection. In some
cases the star was clearly an eclipsing binary; were likely
contaminated by instrumental effects; or showed repeating features
more like outbursts and these were put to one side. Additionally, in
some cases the LS Periodogram had identified half the true period.

There were 730 stars which showed significant variability on a period
$<$1 d. Each lightcurve was further examined by eye and some stars
were removed since the significance of periodic variability was rather
low, leaving 609 stars in our full sample. The shortest period
identified was 1.3 hrs. This highlights the importance of having high
cadence data in identifying UFRs. We also selected a magnitude limited
sample with $T_{mag}<$14.0 mag which consists of 472 stars.

\begin{figure*}
    \centering
    \includegraphics[width = 0.97\textwidth]{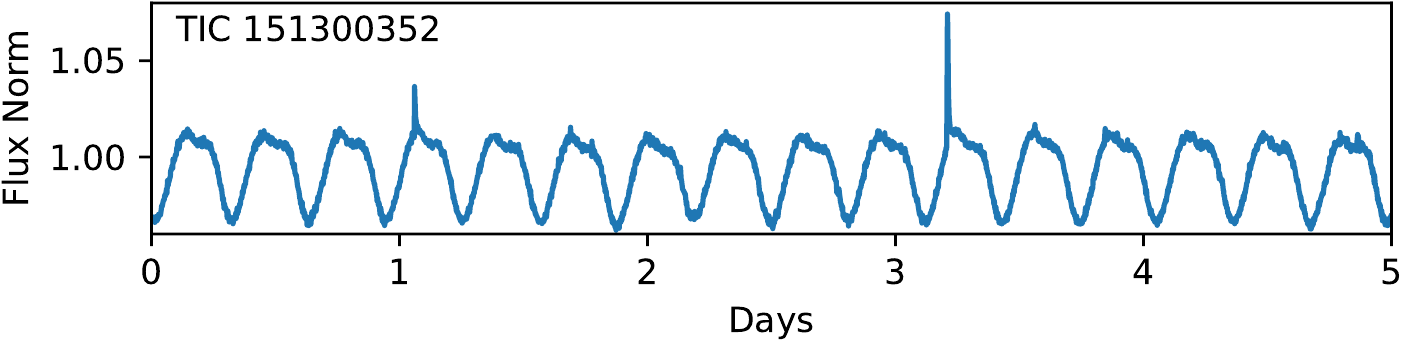}
    \caption{The first five days of the lightcurve of star TIC
      151300352 which has a rotation period of 0.31 d and shows ten
      flares in total although only two here.}
    \label{151300352}
\end{figure*}

\begin{table}
\begin{center}
\begin{tabular}{lr}
\hline
Stars & \\
\hline
247899 & lightcurves in Sectors 1--13 with 2 min cadence;\\
128292 & Unique stars in Sectors 1--13;\\
120439 & Those which have Gaia DR2 data;\\
13836  & Stars close to lower main sequence;\\
12257  & Brighter than $T_{mag}$=15.0 mag;\\
9887  & Brighter than $T_{mag}$=14.0 mag;\\
3613 & $T_{mag}<$15.0 mag and log$_{10}$(FAP)$<$-50\\
730 & LS Period $<$1 d;\\
609 & Likely low mass stars with $P<1$ d;\\
288 & Number of stars which showed at least 1 flare.\\
\hline
\end{tabular}
\end{center}
\caption{The number of stars which were filtered to identify those low
  mass stars which had periodic variability due to short period
  rotation. The more negative the False Alarm Probability (FAP), the
  more likely the periodic modulation is astrophysical and not due to
  noise and systematic variations.}
\label{table1}
\end{table}

\section{Results}
\label{results}

Upon a visual inspection of the lightcurves it was clear that some
stars showed flares, whilst others showed none. As an example of one
star which did show flares, we show the first five days of the light
curve of TIC 151300352 in Figure \ref{151300352} which has a period of
0.31 d and shows 10 flares in a lightcurve of duration 17.4 d
(some data was lost to quality issues).

\footnotetext{\url{http://www.pas.rochester.edu/~emamajek/EEM\_dwarf\_UBVIJHK\_colors\_Teff.txt}}

To help identify flares, we first removed the signature of the
rotation period using {\tt kepflatten} which is part of the {\tt PyKE}
suite of tools\footnote{\url{https://pyke.keplerscience.org}}, on a
sector-by-sector basis. To search for and characterise flares we used
the {\tt
  Altaipony}\footnote{\url{https://altaipony.readthedocs.io/en/latest}}
suite of python based software which is an update of the {\tt
  Appaloosa} \citep{Davenport2016} suite of software. After trial and
error, we selected data points which were 2.5$\sigma$ brighter than
the local mean (higher detection thresholds missed events which were
visible by eye). However, to be classed as a flare, two or more
consecutive flagged points were required. For each flare, the start
time, duration and equivalent duration were determined. Our sample
showed that 288 out of 609 stars showed at least one flare. For our
brighter sample ($T_{mag}<$14.0 mag), we found that 254 out of 472
stars showed at least one flare.

Using our full sample, we compared the brightness of stars showing at
least one flare and showing no flare activity, finding the
distribution of flare active stars is brighter than inactive
stars. This indicates that, unsurprisingly, we are likely to miss
flares from fainter stars. The distributions were similar for stars
brighter than $T_{mag}$=14.0, which is why we also have a bright
sample. At this point we cross matched the bright sample with {\tt
  SIMBAD} and found five stars which showed evidence they were
binaries -- i.e. their spectral type showed a combination of two
spectral types. This reduced the bright sample to 467 stars which we
use for the remainder of our study.

\begin{table*}
  \begin{tabular}{lrrrrrcrrrrrr}  
\hline
    TIC        & RA      &  DEC     & Tmag  & Dist& $M_{G}$ & ($BP-RP$) & $P_{rot}$ & Duration & No. Flares & Flares/day & $L$ & Mean L \\
             & (J2000) & (J2000)  &       & (pc)&           &       & (d)       & (d)      &            &            & (Solar) & (erg) \\
\hline
70775324  & 1.21206 & -30.49875 & 13.75 & 242.6 &  7.8 & 1.77 & 0.2568 & 25.44 & 1 & 0.0393 & 0.074 & 6.6E33 \\
70788006  & 1.74063 & -30.15476 & 13.89 & 237.8 &  8.0 & 1.80 & 0.9450 & 25.44 & 0 & 0.0    & 0.072 &  0.0 \\
70788050  & 1.90691 & -29.99631 & 13.75 & 241.8 &  7.8 & 1.79 & 0.3430 & 25.44 & 0 & 0.0    & 0.080 & 0.0 \\
155896947 & 4.15716 & -52.69336 & 12.88 &  22.8 & 12.6 & 3.34 & 0.1786 & 25.43 & 1 & 0.0393 & 0.001 & 2.9E32 \\
469939293 & 4.44109 & -52.36981 & 13.11 &  82.3 &  9.8 & 2.56 & 0.1919 & 25.44 & 0 & 0.0    & 0.020 & 0.0 \\
425937008 & 5.33499 & -60.92048 & 13.57 &  45.8 & 11.7 & 3.18 & 0.4608 & 50.59 & 0 & 0.0    & 0.003 &  0.0\\
40047077  & 6.46240 & -9.961091 & 11.34 &  32.5 & 10.1 & 2.70 & 0.8888 & 18.68 & 9 & 0.4819 & 0.011 & 8.2E32 \\
246855533 & 6.63975 & -24.20910 & 13.12 &  42.0 & 11.4 & 2.93 & 0.6652 & 25.44 & 2 & 0.0786 & 0.004 & 2.5E32 \\
246891813 & 8.90377 & -24.11765 & 12.80 &  51.5 & 10.6 & 2.77 & 0.7136 & 18.68 & 6 & 0.3212 & 0.006 & 9.4E32 \\
251879914 & 9.93474 & -31.66596 & 12.88 &  73.7 &  9.8 & 2.51 & 0.8840 & 25.43 & 1 & 0.0393 & 0.015 & 1.4E33 \\
  \hline
  \end{tabular}
  \caption{A summary of the stars in our sample. We show the stars TIC
    number; RA and DEC; the $T_{mag}$; distance determined using Gaia
    DR2; absolute Gaia $G$ mag determined from the distance and mean
    Gaia $G$ mag; the rotation period determined using the LS
    periodogram; the duration of the lightcurve; the number of flares
    detected; the number of flares per day; the stars quiescent
    luminosity in Solar units and the mean luminosity of the
    flares. Here we show only ten stars for illustrative purposes and
    the full table is available in fits format in the on-line
    version.}
  \label{flares}
  \end{table*}

To determine the quiescent luminosity of the stars in our sample we
again used the Gaia $G$ band absolute magnitude and equation 4 from
\citet{Andrae2018}:

\begin{equation}
-2.5\log_{10} L = \mg + \bcg(\teff) - \mbolsun
\label{luminosity}
\end{equation}

where $L$ is the quiescent stellar luminosity in units of \lsun
(3.83$\times10^{33}$ erg/s); \mg\ is the absolute magnitude of the
star in the $G$ band; $\bcg(\teff)$ is a bolometric correction
(determined using equation 7 of \citet{Andrae2018}), and $\mbolsun =
4.74$\,mag is the Solar bolometric magnitude. Since almost all of our
sample is within 220 pc we did not take into account reddening since
it likely to be low.

To determine the luminosity of the flares, we take the equivalent
duration of the flare, calculated using {\tt Altaipony}, and multiply
this by the stellar quiescent luminosity, calculated using equation
\ref{luminosity}. The mean energy of the flares was
$1.7\times10^{33}$~erg and the flare with the highest energy was
$2.2\times10^{35}$~erg, from TIC 100481123 with $P_{rot}$=0.84 d. The
lowest energy flare was $2\times10^{31}$~erg (in solar terms this
would equate to an X2 class flare), whilst the mean of the lowest
energy flares was 8$\times10^{32}$~erg. In Table \ref{flares} we give
an overview of the flare characteristics of each star while in Figure
\ref{gaia-hrd} we show the location in the Gaia HRD where the symbol
size indicates the flare rate.

We split up the bright sample into stars which showed at least one
flare per 17.38 d (representing the shortest duration of a light
curve in our sample) and no flares per 17.38 d. We find no clear
evidence for a difference in the distribution of stars in the Gaia HR
Diagram.

\begin{figure}
\begin{center}
  \includegraphics[width=0.47\textwidth]{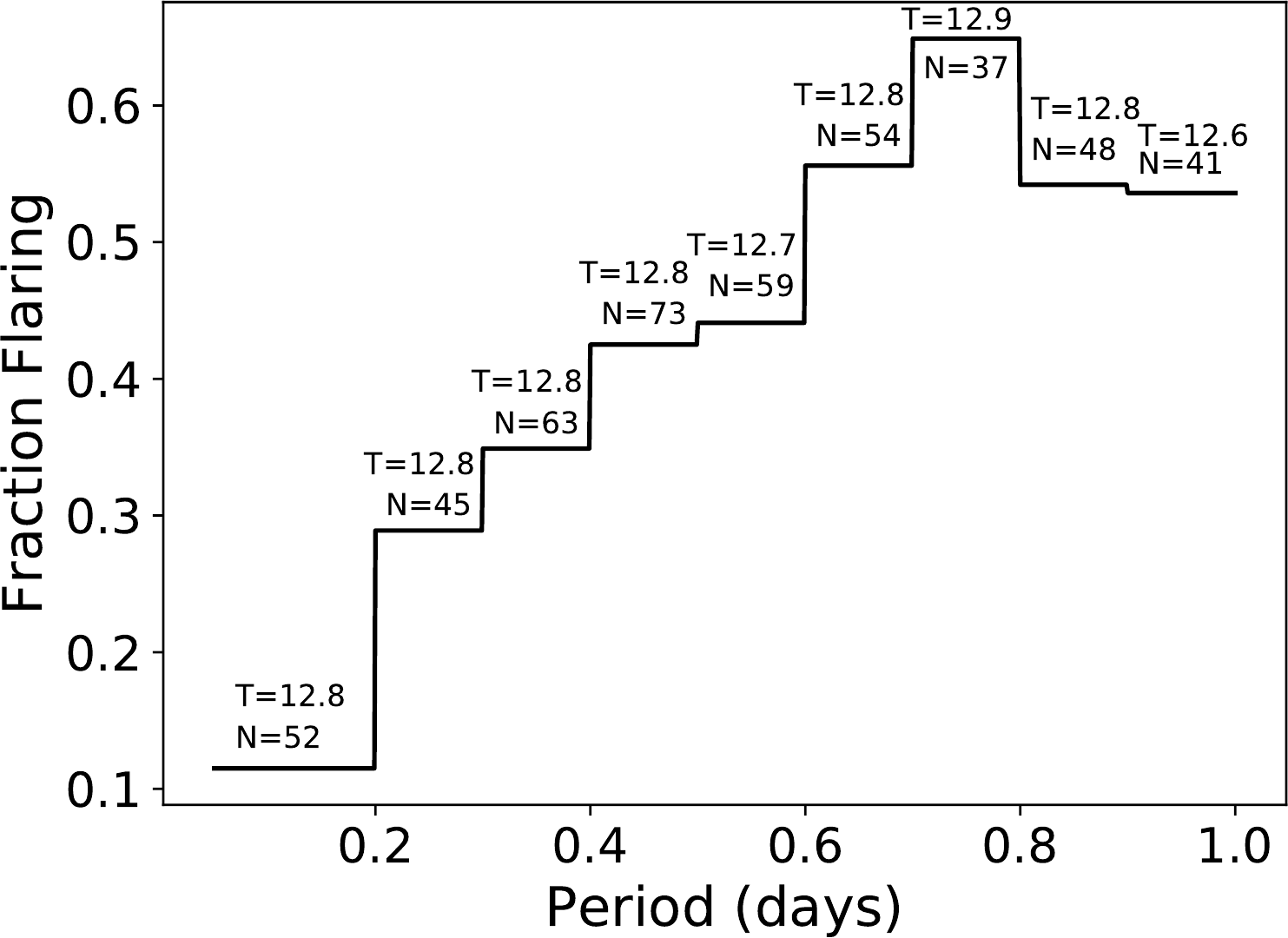}
  \caption{Using our sample of stars brighter than $T_{mag}$=14 we
    show the fraction of stars in each period bin which show more than
    0.044 flares per day (see text for details). The numbers of stars
    in each bin is shown above or below the histogram. There is a
    clear decline in the fraction of stars as the period gets shorter
    with a rapid drop at periods $<$0.2 d. We also show the mean
      mag of stars in each bin: there is no significant change over
      the period range.}
    \label{fracflareperiod}
    \end{center}
\end{figure}

We also determined the fraction of stars in our bright sample which
showed at least one flare over the median duration of the lightcurves
(22.7 d), over various period ranges and show this as a histogram in
Figure \ref{fracflareperiod}. There is a clear decline in the fraction
of stars which show flaring activity as the period gets shorter, with a
rapid decline at periods of $<$0.2 d. There is no clear
difference between the energy of those flares which are detected over
this period range. At periods $>$0.4 d, the average fraction of
stars which show at least one flare per 22.7 d is 51 percent, whilst
in the $<$0.2 d bin, the fraction is 11.5 percent. This has a
probability of being due to chance of $<10^{-5}$ which is significant
at a level of $>5\sigma$. There is no significant difference in the
median duration of the lightcurves in each period bin. We also
determined the fraction of stars showing flares at a rate of one per
17.38 d, the shortest duration of the lightcurves, and find the
same result.

In Figure \ref{fracflareperiod} we also show the mean $T_{mag}$ for
each bin. There is no significant difference in the mean brightness of
stars over the period range and, therefore, we can exclude the
possibility that the shortest period stars show fewer flares simply
because they are fainter. We also determined the mean amplitude of the
lightcurves in each bin and find no correlation with the period bin. A
further search was made for a variation with both $(BP-RP)$ and the
mean Galactic latitude of the samples in each bin. We find that the
sample with $P<$0.2 d is slightly bluer ($BP-RP\sim2.5$) than stars
with $P>0.2$ d ($BP-RP\sim2.8$), and that the mean Galactic latitude of
the sample with $P<$0.2 d is $|b|\sim18.5^{\circ}$ compared with
$|b|\sim30^{\circ}$ for the $P>$0.2 d sample. We return to these
findings in \S \ref{discussion}.

\section{Comparison with the Kepler field}
\label{kepcomp}

We took all the stars observed using {\kep} in short cadence mode in
the original field and identified those which are likely single low
mass stars using the same criteria as described in \S \ref{tess}. As a
result, we found 73 unique stars. In contrast, we have found nearly
10,000 stars brighter than $T_{mag}$=14 close to the lower main
sequence in the southern ecliptic hemisphere which were observed using
{\tess} in 2 min cadence. This is more than two orders of magnitude
more than were observed using {\kep}. We searched the {\kep}
lightcurves for a periodic signal and found three which were $<$1
d. One, KIC 3730067, is an eclipsing binary with a period of 0.147 d,
while KIC 10063343 has a period of 0.333 d and KIC 9726699 (also known
as GJ 1243) has a period of 0.593 d.

With {\tess} starting observations of the northern ecliptic hemisphere
in July 2019, we have the opportunity to compare the lightcurves of
specific stars made using {\kep} and {\tess}. KIC 10063343 does not
appear to have been observed so far using {\tess}. However, KIC
9726699, (TIC 273589987) has been observed so far in sectors 14 and
15. Indeed, KIC 9726699 was found to be a UFR with a rotation period
of 0.593 d by \citet{Ramsay2013} which showed flares every few hours.

We were able to make a direct comparison of the lightcurve of GJ 1243
($T_{mag}$=10.3 and $G$=11.6) derived using {\kep} and {\tess} using
the same methods described in \S \ref{tess}. There is a gap of seven
years between the end of the {\kep} observations and the start of the
{\tess} observations. Using the lightcurves which have the signature
of the rotation period removed (and using a 5$\sigma$ clipping) we
find that the rms of the {\tess} lightcurves are a factor of 4.6
higher compared to the {\kep} lightcurves. This is entirely expected
as the collecting area of the {\kep} mirror is much greater than the
aperture of the {\tess} telescopes.

\begin{figure}
\hspace{-3mm}
  \includegraphics[width=9cm]{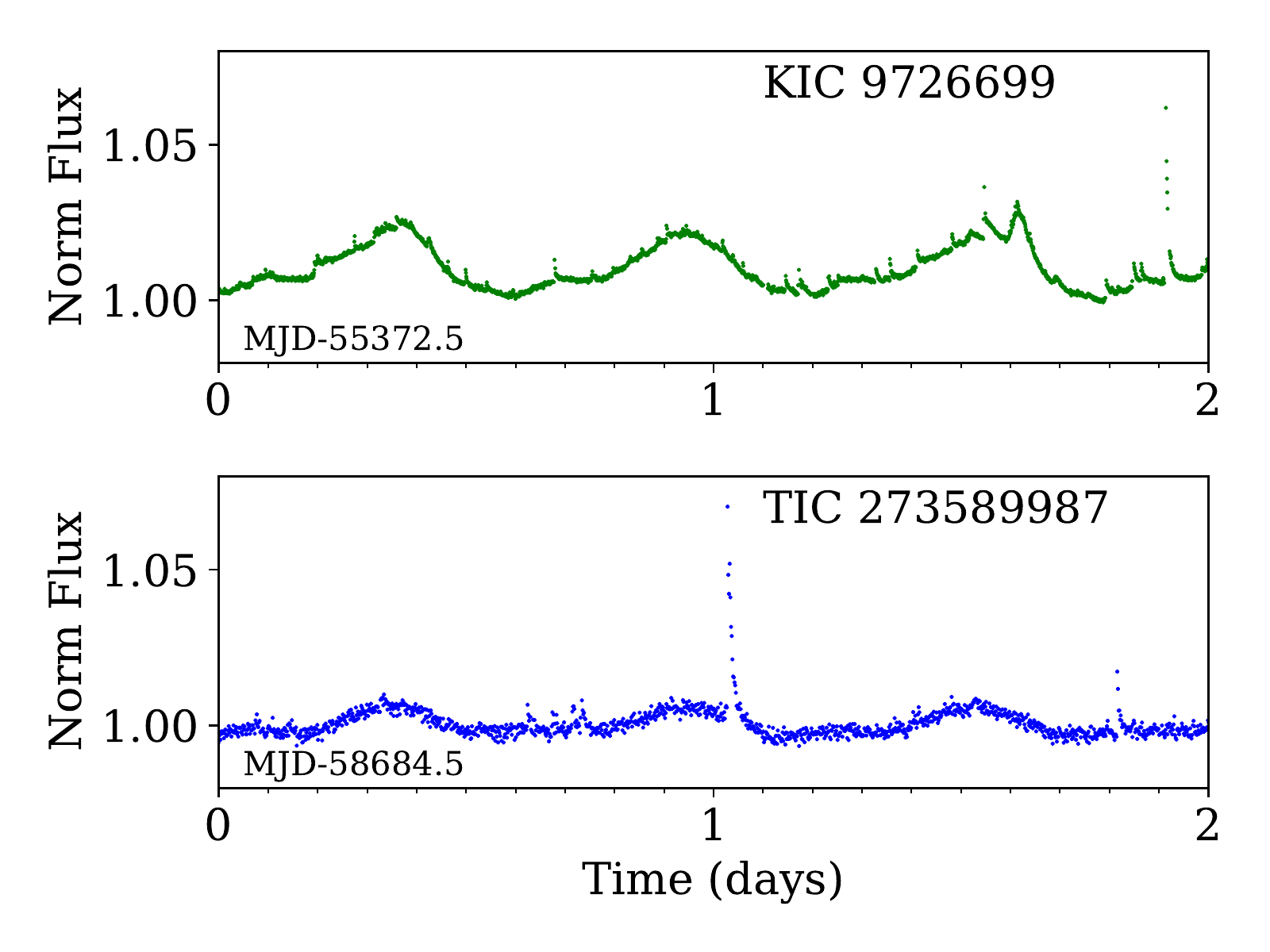}\hfill
\vspace{-8mm}
    \caption{Observations of the flare star GJ 1243 made using {\kep}
      (top panel) and {\tess} (lower panel). With the larger effective
      aperture of {\kep} more lower energy flares are detected
      compared to {\tess}.}
    \label{kepler-tess}
\end{figure}

We identified all the flares in both the {\kep} and {\tess} light
curves and show a two day section from both lightcurves in Figure
\ref{kepler-tess}. In the {\kep} lightcurve there are more short
duration, low amplitude flares than the {\tess} lightcurve. The
amplitude of the modulation is also higher in the {\kep} lightcurve
which is likely due to the larger pixel size of the {\tess} detectors
(21$^{''}$ per pixel). This will result in the dilution of the
  varying signal of the target if there are other stars spatially
  located in or close to the pixel. This is especially true at low
  Galactic latitudes, for example, GJ 1243 with $b=+10^{\circ}$.

In \citet{Ramsay2013} we assumed the quiescent luminosity for GJ 1243
was typical of an M4V star ($3.6\times10^{30}$ erg/s). However, as can
be seen from Figure \ref{gaia-hrd} there is quite a large spread in
absolute magnitude of stars M3--M4V. Applying the same methods as
outlined in \S \ref{tess} we estimate that the quiescent luminosity of
GJ 1243 is $1.7\times10^{31}$ erg/s -- a factor of 5 greater than
assumed before. As expected from the difference in the rms of the
lightcurves, we find that {\kep} has a mean flare energy of 4.5 lower
than that of {\tess} and an order of magnitude difference in the
lowest energy flare detected (see Table \ref{kep-tess-energy}). Although
we cannot rule out that some of this difference is due to a change in
the activity levels between the two epochs, we believe this is simply
due to the {\kep} lightcurves having a much lower rms because of the
size of the {\kep} mirror. {\tess} is, as expected, unable to detect
the lower energy flares that were detectable using {\kep}.

\begin{table}
\begin{center}
\begin{tabular}{lrrr}
  \hline
  & Mean Energy & Lowest Energy & Highest Energy \\
  & (ergs) & (ergs) & (ergs) \\
  \hline
  {\tess} & $1.5\times10^{32}$ & $3.0\times10^{31}$ & $6.2\times10^{32}$ \\
  {\kep} & $3.3\times10^{31}$  & $2.9\times10^{30}$ & $1.1\times10^{33}$ \\
\hline
\end{tabular}
\end{center}
\caption{A comparison of the mean, lowest and highest energy flares
  detected from GJ 1243 using 3 years of {\kep} data and 2 months of
  {\tess} data. }
\label{kep-tess-energy}
\end{table}

\section{Discussion}
\label{discussion}

Our sample of stars covers the lower main sequence from K9 to M6 and
have identified UFRs all along its length. We find that stars which
are flare active are found all along the sequence. However, we also
find evidence that stars with $P<$0.2 d show significantly fewer
flares than those stars with periods in the range 0.3--1.0 d.  This is
consistent with the finding of \citet{Gunther2020} who used 2 min
cadence data taken from the first two months of the {\tess} mission,
and found there was a `tentative' decrease in the flare rate for stars
with $P<$ 0.3 d. Rapidly rotating low mass stars are expected to
produce increased levels of activity, including flares, which is
strongly related to their dynamo mechanism
\citep{Hartmann1987,Maggio1987}. Our finding, and that of
\citet{Gunther2020}, that stars with apparently the shortest rotation
period show the fewer flares is therefore a surprise.

We first discuss how stable magnetic fields are on UFRs. One {\sl
  active} UFR, V374 Peg ($P$=0.44 d, M3.5), has received a lot of
attention \citep[e.g.][]{Vida2016}. They showed that the spot
configuration was stable over about 16 years, confirming previous
indications of a very stable magnetic field. Additionally, they
observed frequent flaring, with the stronger flares seeming to be more
concentrated around the phase where the lightcurve indicates a smaller
active region. \citet{Morin2008} carried out a detailed modelling of
the objects magnetic topology finding it had a large-scale magnetic
topology, mostly axi-symmetric poloidal, but with the configuration
sheared by very weak differential rotation. This suggests that a few
low-contrast spots are present on the photosphere of V374 Peg,
covering only $\sim$2 percent. On the other hand, more recently,
\citet{Lanzafame2019}, found evidence using multi-epoch data from Gaia
DR2, that stars younger than 600 Myr, especially UFRs, can quite
rapidly change their amplitude, indicating a comparable change in
their spot coverage and hence activity level.

\citet{Reiners2012,Reiners2014} showed that fast rotating stars could
exhibit a strong H$\alpha$ emission line which was saturated as the
stars rotation rate increased. One of the stars in our study with
$P<$0.2 d, TIC 260504446, was detected using {\sl XMM-Newton} and has
an observed luminosity of $L_{X}\sim8\times10^{28}$ erg/s using the
{\sl Gaia} parallax. Assuming an optical luminosity of
2.6$\times10^{31}$ erg/s (c.f. \S \ref{results}), we determine
log$(L_{X}/L_{bol})\sim10^{-3}$ for TIC 260504446, implying it falls
into the saturated X-ray regime \citep{Reiners2014}. The work of
\citet{Reiners2012} indicates that $L_{H\alpha}$ could be as high as
8$\times10^{27}$ erg/s, or 1/10 of the X-ray luminosity \citep[see
  also][]{Jeffries2011}. We predict that this flare active star, with
on average one flare per 0.08 d seen in the {\tess} data, will show
strong H$\alpha$ emission.

As stars age they lose angular momentum through processes such as
stellar winds and, therefore, slow down \citep[e.g.][and references
  therein]{BarnesKim2010}. Determining the rotation period of stars in
open clusters of known ages gives direct information on how age
affects the rotation rate. For instance, using data from {\sl K2},
\citet{Rebull2016,Rebull2017} determined the period of stars in M45
(125 Myr) and M44 (630 Myr). They found that of those stars with a
known period, the percentage of stars $P<$1 d is 57.1 percent and 20.6
percent for M45 and M44 respectively, with these numbers declining to
3.2 percent and 2.1 percent respectively for $P<0.2$ d. Although there
are inevitably differences in the biases between the {\sl K2} sample
and our {\sl TESS} sample, we find 33.1 percent and 2.9 percent have
$P<1$ d and $P<0.2$ d respectively -- i.e. between the values for M45
and M44. In Figure \ref{gaia-hrd} we show age tracks which indicate
that the stars in our sample are all older than $\sim$30 Myr and many
appear older than $\sim$100 Myr, which is consistent with the
comparison with M45 and M45. Based on their position in the Gaia HRD,
we are unable to determine if age could be a dominant factor in the
apparent lack of flare activity in our study.

We now consider whether the source of the short periodic signal could
be another star located spatially close to the target star. This is
because of the pixel scale of the TESS detectors (21 arcsec per pixel)
which gives a 90 percent encircled radius of 42 arcsec
\citep{Ricker2015}. Crowding is especially an issue near the Galactic
plane. Indeed, of those stars with $P<$0.2 d, 56 percent are within
15$^{\circ}$ of the Galactic plane, indicating crowding could be an
issue. However, high cadence photometric surveys such as OmegaWhite
\citep{Macfarlane2015,Toma2016} allow us to estimate how many variable
stars with $P>$ 1.3 hrs are present at low Galactic latitude. Based on
this 400 square degree survey, we estimate that for stars brighter
than $g$=16, we expect less than one variable star with a period
greater than 1.3 hrs to be a chance alignment with our targets. We,
therefore, do not think it likely that the variability seen in our
targets is due to a chance alignment with a spatially nearby variable
star.

Could the rotation periods we detect, in fact, not be the signature of the
rotation period of the star, but rather the signature of a binary orbital period?
Although we have restricted our search of stars close to the lower
main sequence where we expect most stars to be late type single stars,
we know that interacting binaries such as cataclysmic variables (CVs)
and white dwarf - red dwarf pairs can also lie in this part of the
HRD. Further, we expect that around 1/4 of early M type stars are in
binaries \citep[e.g.][and references therein]{Cortes-Contreras2017}.

Theoretical work, such as \citet{Counselman1973}, and observations
outlined and referred to by \citet{Fleming2019}, indicate that
for such short period binaries the stars are very likely synchronised
with the orbital period. If the flare inactive UFRs are binaries, then
the stars are likely to have increased their rotation rate as the
orbit separation has shrunk, due to angular momentum loss over
time. The stars rotation period could, therefore, be much shorter than
expected for its age if it were not in a binary system. A series of
medium resolution spectra would be required to test whether these
flare inactive stars are in fact members of binaries.

Perhaps an obvious reason for our finding is that the flare inactive
short period stars do show flares, but they are not seen in the
{\tess} data. We showed in \S \ref{kepcomp} that for observations of
the same star, {\kep} was sensitive to flares with an energy
$3\times10^{30}$~erg while {\tess} was sensitive to flares
$3\times10^{31}$~erg. A related explanation is that they do show
flares but are less energetic than stars with slower rotation
velocities. Dedicated observations of some of these stars using larger
telescopes and high cadence multi-band photometry extending down to
the $U$ band, where much of the flare energy is thought to be emitted,
might be able to shed light on this.

A related solution could be, if a stars activity is saturated then the
chromosphere of the M dwarfs becomes a very efficient radiator in blue
continuum wavelengths. This could be the result of increased
micro-flaring \citep[see][]{Houdebine1996}. These flare inactive UFRs
may, therefore, show an enhancement in the blue optical band, which
would be apparent in the $U-B$ color index, but not in $B-V$. This can
be tested via spectroscopic observations and $U$ band flare
monitoring.

In \S \ref{results} we noted that the stars in our sample with $P<0.2$
d were slightly closer on average to the Galactic plane
($|b|\sim18.5^{\circ}$) compared to those stars with $P>0.2$ d
($|b|\sim30^{\circ}$). This would dilute the target signal and reduce
the apparent amplitude of any rotational modulation or flare
amplitude. This is confirmed when we compare the number of stars in
Gaia DR2 which are within 42~arcsec and have a $G$ mag within at least
1.5 mag of the target. There are around 10 percent more spatially
nearby stars in the $P<0.2$ d sample. However, this is not large
enough to explain the drop in the fraction of flare active stars in
the $P<0.2$ d sample. The fact that the amplitude of the rotational
modulation is not correlated with period does not support this
either. We also noted in \S \ref{results} that the $P<0.2$ d sample is
slightly bluer ($BP-RP\sim2.5$) compared with stars with $P>0.2$ d
($BP-RP\sim2.8$). Given they are also closer to the Galactic plane, this may
indicate that they are biased towards younger stars. In which case we
would expect them to be more active not less active. Could this
indicate that we are seeing the enhanced emission due to increased
magnetic activity at bluer wavelengths? \citet{Houdebine1997}
reported a blue excess in active dMe stars that increased with
activity level which was about three times larger than expected from
calculations for a given atmospheric pressure.

\section{Conclusions}

We have identified a sample of low mass stars which lie close to the
main sequence which show a periodic modulation with $P<$1 d. We find
that the fraction of stars which are flare active declines at shorter
periods, with a significant drop at $P<$0.2 d. Since we associate
rapid rotation with strong magnetic fields which is likely to drive
optical flares, this is a surprise. Therefore, we have explored
several potential reasons for this. We demonstrate that it is unlikely
to be due to spatially close variable stars being the source of the
modulation. We find that the sample with $P<$0.2 d is slightly bluer
and closer on average to the Galactic plane than those with $P>0.2$
d. We explore the effect that spatially nearby stars may dilute the
target stars rotational amplitude. This would reduce the number of
detected flares but conclude it is unlikely to be the sole reason for
our finding. If the flare inactive stars with $P<$0.2 d were binary
stars, then the stellar components are likely to have increased their
rotation rate as their orbital separation shrinks due to angular
momentum loss over time. They would be rotating faster than if they
were single stars and therefore their rotation period would not
reflect their age. Several medium resolution spectra are needed to
search for evidence of binarity and to determine the flux of lines
such as H$\alpha$. Another possibility is that these stars show low
energy flares, perhaps at bluer wavelengths which would not be
detected using {\tess}. High cadence photometry of stars with $P<$0.2
d, especially in the $U$ band, is encouraged.

\section{Acknowledgments}

We thank the anonymous referee for a helpful report which made several
very useful suggestions. This paper includes data collected by the
TESS mission, for which funding is provided by the NASA Explorer
Program. Armagh Observatory \& Planetariun is core funded by the
Northern Ireland Executive through the Dept for Communities. LD
acknowledges funding from an STFC studentship.
 
This work presents results from the European Space Agency (ESA) space
mission {\sl Gaia}. {\sl Gaia} data is being processed by the {\sl
  Gaia} Data Processing and Analysis Consortium (DPAC). Funding for
the DPAC is provided by national institutions, in particular the
institutions participating in the {\sl Gaia} MultiLateral Agreement
(MLA). The Gaia mission website is
\url{https://www.cosmos.esa.int/gaia}. The Gaia archive website is
\url{https://archives.esac.esa.int/gaia}.

Data Availablity: The {\tess} data is available from the NASA MAST
portal.

\end{document}